\RequirePackage{amsmath}
\documentclass[cmp,referee]{svjour}  

\journalname{Communications in Mathematical Physics}

\usepackage{lineno}
\usepackage[hidelinks]{hyperref}
\usepackage{color}
\usepackage{amsfonts,amssymb}
\usepackage{subfigure, float, graphicx}
\usepackage[lmargin=1in,rmargin=1in]{geometry}
\graphicspath{{Figs/}}
\usepackage[title,titletoc,toc]{appendix}

\modulolinenumbers[5]
\newcommand{\tr}{\ensuremath \textnormal{tr}}

\begin{document}

\title{{Energy Variation of Soft Matter Interfaces}}
\titlerunning{Energy Variation of Soft Matter Interfaces}

\author{Prerna Gera and David Salac\inst{1}}
\institute{Department of Mechanical and Aerospace Engineering, University at Buffalo,
	Buffalo, New York 14260-4400.\\ \email{davidsal@buffalo.ed}}
\authorrunning{P. Gera \& D. Salac}

\maketitle

\begin{abstract}
  The variation of energies associated with soft matter interfaces 
  where surface inhomogeneities are present.
  These energies include the total bending and splay energy, the variable surface
  tension energy, a coupling energy between the total curvature and an underlying 
  surface concentration field, the energy due to an external field, and a phase segregation
  energy. When considering these energies the variation of material properties 
  such a bending rigidity are taken into account, which results in 
  more general variation expressions.
  These variations can be used to determine the equilibrium 
	interface and concentration configuration or to determine the 
	driving forces for non-equilibrium situations. 
	While the focus of this work are energies associated with multicomponent
	vesicles, it can easily be extended to any soft matter interface.
\end{abstract}

\section{Introduction} 
Soft matter interfaces play a critical role in a large
number of material systems. For example, additives 
used in enhanced oil recovery techniques induce low interfacial tension between water and
oil which helps in oil displacement and final recovery~\cite{salager2012emulsion,sagis2011dynamic}.
By controlling the types of surfactants
on the interface the emulsions can be inverted~\cite{perazzo2015phase}.
The interplay between the interface and the surrounding fluids 
can be used to make skin care products, where 
nano-emulsions or polymer
thickening agents are heavily used~\cite{perazzo2015phase}.


Soft matter interfaces also play a crucial role in the behavior of biological
systems such as red blood cells. These interfaces
play a vital role since they not only acts as a protective barrier to the cell interior, 
but they also act as the medium of communication with the
environment outside. The composition of the biological membrane also has a direct impact on the
fundamental biological processes such as signal transduction, trafficking and
sorting processes~\cite{simons2000lipid,simons2004model,mukherjee2004membrane}.

These biological membranes are constituted of various kinds of lipids and cholesterol. As
these molecules move freely in the plane of the membrane they often combine
to form domains that are energetically more stable than the rest of the
membrane. Mechanical properties, such as bending rigidity and membrane spontaneous
curvature, can vary depending on the local membrane composition and molecular arrangement. 
This variation of properties influences not only the the segregation and 
coarsening processes of the membrane domains, but also the surrounding fluid.
The coupling between the composition of the membrane and changes in it's morphology is
therefore interesting and significant. 

Pioneering work on biological soft matter interfaces have been done by Canham~\cite{canham1970minimum},
Helfrich,~\cite{helfrich1973elastic}, and Evans~\cite{evans1974bending}. They
individually studied the mechanics of the membrane and presented the free energy
functional which depends purely on geometric quantities. 
The variation of this functional was taken to determine the Euler-Lagrange,
or ``Shape" equation~\cite{zhong1987instability,zhong1989bending}. This gave rise to
numerous theoretical 
investigations~\cite{biben2005phase,biben2003tumbling,du2007analysis,du2006simulating,jamet2007towards}, however, 
most of these focused on a homogeneous
membranes. 

In this work the variation
with respect to changes of the interface location
and changes of a surface concentration field is considered.
These variations can be used to determine the equilibrium 
interface and concentration configuration or to determine the 
driving forces for non-equilibrium situations. 
While the focus of this work are energies associated with multicomponent
vesicles, it can easily be extended to any soft matter interface including
droplets~\cite{chatterjee2006droplet}, bubbles~\cite{stefan1999surfactant,davis1966influence}, 
fiber-laden membranes~\cite{neville1993biology}, and
biopolymers~\cite{sagis2011dynamic}.

The energies considered here include the membrane total and splay bending energies,
surface tension, the coupling between the total curvature and concentration field,
a generic external field, such as an electric or magnetic field, and a phase energy consisting of gradient and mixing terms.
The methodology used here is based on the  work of Napoli and
Vergori~\cite{Napoli2010}. Unlike the prior work, it is assumed that all
material parameters, such as total bending rigidity and spontaneous curvature,
depend on the underlying concentration field. This results in more general 
expressions which are valid for a larger number of systems.

The outline of the paper is as follows. The mathematical framework is outline in 
Section \ref{sec:framework}. Here the general expression for variation of the interface free
energy with respect to changes of the interface location and surface concentration field will be presented.
Specific energy cases are shown in Section \ref{sec:specific_cases}, where each free energy 
is considered separately. A brief discussion follows in Section \ref{sec:discussion}. To provide 
additional clarity, Appendices \ref{sec:surfaceCalculus} and \ref{sec:variationL} provide information
about various surface calculus and variational identities used in Sections \ref{sec:framework} and \ref{sec:specific_cases}.

\section{Mathematical Framework}
\label{sec:framework}
The mathematical framework used here is based on the work of Napoli and Vergori~\cite{Napoli2010}. 
In this prior work a systematic method is developed to obtain the equilibrium equations 
for nematic crystal vesicles. In this section the prior results relevant
to the current work are briefly outlined. The addition of an additional 
energy contribution not considered by Napoli and Vergori is also shown.

Consider a closed interface $\Gamma$ with 
an outward facing unit normal of $\vec{n}$ which separates two fluids.
There could possibly be two components to this interface, with the concentration
given by $q$.
This interface is characterized by the second fundamental form, also called the shape tensor,
given by $\vec{L}=\nabla_s \vec{n}$, where $\nabla_s$ represents the surface 
gradient. This is a symmetric second-order tensor field which only
has components tangent to the interface. It also only has two non-zero
eigenvalues, $c_1$ and $c_2$, which are called the principle curvatures.
Using this definition the total and Gaussian curvature can be defined as
\begin{align}
	H&=c_1+c_2=\tr\vec{L}=\nabla_s \cdot \vec{n}, \\
	K&=c_1c_2=\frac{1}{2}\left[\left(\tr \vec{L}\right)^2-\tr\left(\vec{L}^2\right)\right],
\end{align}
respectively.

The free energy functional for the interface is defined on the closed surface
$\Gamma$ as
\begin{align}
	W[\Gamma]=\int_\Gamma w( \vec{r},\vec{n},\vec{L},q,\nabla_s q)\;dA,
\end{align}
where $w( \vec{r},\vec{n},\vec{L},q,\nabla_s q)$ is the free energy density
which may depend on surface quantities $\vec{n}$, $\vec{L}$, $q$, and $\nabla_s q$
and the position of the interface, $\vec{r}$. 

To calculate the first variation of the free energy, assume that the interface 
$\Gamma$ undergoes an infinitesimal virtual displacement, 
\begin{align}
	\vec{r}'=\vec{r}+\epsilon \vec{u},\label{rvar}
\end{align}
where $\epsilon$ is a small positive parameter and
$\vec{u}$ is a virtual displacement field. The prime denotes the quantities and
operators relative to the virtually deformed configuration. The variation of a quantity
is defined as 
\begin{equation}
	\delta(\cdot)=\lim_{\epsilon\to0}\dfrac{(\cdot)'-(\cdot)}{\epsilon},
\end{equation}
where $(\cdot)$ denotes any quantity defined on $\Gamma$.
Using the transport
theorem, the variation of the free energy can be written as~\cite{Napoli2010}
\begin{align}
	\delta W[\Gamma]=\int_\Gamma (\delta w + w \nabla_s \cdot \vec{u}) \;dA,
\label{delta_W}
\end{align}
where
\begin{align} \delta w=\frac{\partial w}{\partial \Gamma}\delta\Gamma+\frac{\partial w}{\partial \vec{n}} \cdot \delta\vec{n}
	+\frac{\partial w}{\partial \vec{L}} \cdot \delta \vec{L} +\frac{\partial
	w}{\partial q}\delta q + \frac{\partial w}{\partial (\nabla_s q)}\cdot \delta
	\nabla_s q.\label{varTotEnergy} 
\end{align}
The component $\left(\partial w/\partial \Gamma\right)\delta\Gamma$
provides the change of the free energy when the interface undergoes bulk shape changes while the others capture changes
for interface-only quantities.

The individual components are 
\begin{align}
	\delta \Gamma & = \delta \vec{r} \cdot \vec{n}=\vec{u}\cdot\vec{n}, \label{delta_gamma}\\
	\delta \vec{n} &= -(\nabla_s \vec{u})^T \vec{n} \label{delta_n},\\
	\delta \vec{L} &= \vec{L}(\nabla_s \vec{u})^T \vec{n} \otimes \vec{n} -
		\nabla_s[(\nabla_s\vec{u})^T\vec{n}]-\vec{L}(\nabla_s \vec{u}), \label{delta_sgn}\\
	\delta(\nabla_s q) &= \nabla_s \delta q +[(\nabla_s \vec{u})^T\vec{n} \cdot
		(\nabla_s q)]\vec{n} -(\nabla_s \vec{u})^{T} \nabla_s q. \label{delta_q}
\end{align}
Forms for $\delta \vec{n}$, $\delta(\nabla_s q)$, and $\delta \vec{L}$ have been shown previously~\cite{Napoli2010}.
As the definition of $\vec{L}$ presented here differs from Napoli and Vergori, 
the derivation of $\delta \vec{L}$ has been included in Appendix \ref{sec:variationL}. 

Introduce the conjugate variables $\beta$, $\vec{\Lambda}$, $a$, $\vec{b}$, and $f$,
\begin{align}
	\beta=\frac{\partial w}{\partial \vec{n}},\qquad
	\vec{\Lambda}=\frac{\partial w}{\partial \vec{L}}, \qquad
	a=\frac{\partial w}{\partial q}, \qquad
	\vec{b}= \frac{\partial w}{\partial(\nabla_s q)}, \qquad
	f= \frac{\partial w}{\partial \Gamma}=\nabla w\cdot\vec{n},
	\label{defconvar}
\end{align}
where $\nabla w$ only applies to those terms of $w$ with explicit dependence on spatial location $\vec{r}$.
It is then possible to write Eq.~(\ref{varTotEnergy}) as 
\begin{align}
	\delta w = [(\nabla w \cdot \vec{n})\vec{n}]\cdot \vec{u} +\nabla_s \cdot \left\{ [(\nabla_s \vec{u})\vec{\Lambda}_s]^T \vec{n} +b_s
		\delta q \right\}+\vec{\sigma}_E\cdot \nabla_s \vec{u} + (a-\nabla_s \cdot \vec{b}_s)\delta q
		\label{delta_w}
\end{align}
where $\vec{\sigma}_E$ is
\begin{align}
	\vec{\sigma}_E = -\vec{L}\vec{\Lambda}_s-\nabla_s
	q\otimes\vec{b}_s-\vec{n}\otimes\left\{\vec{P}(\beta-\nabla_s\cdot\vec{\Lambda})-\vec{L\Lambda
	n} -(\vec{b}\cdot\vec{n})\nabla_sq\right\}
\end{align}
with
$\vec{\Lambda}_s=\vec{\Lambda}\vec{P}$
and $\vec{b}_s=\vec{P}\vec{b}$.
Using these expressions the variation of the free energy can then be written as
\begin{align}
	\delta W =\int_{\Gamma}[(\nabla w \cdot \vec{n})\vec{n}-\nabla_s \cdot	\vec{\sigma}]\cdot\vec{u}\; dA
	+\int_{\Gamma} [a-\nabla_s\cdot\vec{b}_s ]\delta q\; dA 
	\label{eq:final_varTotEnergy}
\end{align}
where
\begin{align}
	\vec{\sigma}&=w\vec{P}+\vec{\sigma}_E. \label{sigmatot}
\end{align}
From this the variation of the energy due to changes in the interface is
given by 
\begin{align}
	\vec{\mathcal{F}}_\Gamma&=f\;\vec{n} - \nabla_s \cdot \vec{\sigma} \label{GovernFF}
\end{align}
while the variation of the energy due to changes in the concentration field 
is given by
\begin{align}
	\mathcal{F}_q&=a-\nabla_s\cdot\vec{b}_s. \label{GovernSPF}
\end{align}

At equilibrium $\delta W=0$ for arbitrary $\vec{u}$ and $\delta q$ and thus
both Eqs. (\ref{GovernFF}) and (\ref{GovernSPF}) must equal zero.
When not in equilibrium the variations are related to the forces which drive the system to equilibrium.
For example, consider an interface surrounded by fluid where the surface concentration is modeled using the Cahn-Hilliard equation. 
The variation associated with the interface, $\vec{\mathcal{F}}_\Gamma$, 
would be related to the force exerted by the interface on the surrounding fluid 
while the variation with respect to the surface concentration, $\mathcal{F}_q$, would define the 
chemical potential.

\section{Specific Cases}
\label{sec:specific_cases}
Using the framework shown in Section \ref{sec:framework} the resulting variations are derived for the case of an
interface with total and Gaussian curvature energy, variable surface tension, coupling energy between
the surface concentration and the interface curvature, an external field, and
where the surface concentration is described using a phase-field energy form. A typical
system which contains all of these energies would be a vesicle membrane with multiple 
lipid species. By ignoring 
the curvatures energy, it is possible to describe a fluid-fluid or fluid-air interface with a surfactant.

The six energies considered here are:
\begin{align}
	W_b[\Gamma]&=\int_\Gamma{\frac{k_c(q)(H-c_0(q))^2}{2} \;dA},\label{EnIntBending}\\
	W_s[\Gamma]&=\int_\Gamma{k_g(q)K \;dA},\label{EnIntSplay}\\
	W_\gamma[\Gamma]&=\int_\Gamma{\gamma(q) \;dA},\label{EnIntTension}\\
	W_{c}[\Gamma]&=\int_\Gamma{\eta H q \;dA},\label{EnIntCoup}\\
	W_{ext}[\Gamma]&=\int_\Gamma{\frac{m(q)}{2}\vec{F}(\vec{r})\cdot\vec{F}(\vec{r})\;dA},\label{EnIntMag}\\
	W_{q}[\Gamma]&=\int_\Gamma \left[\frac{k_f}{2}\|\nabla_s q\|^2+g(q)\right]\;dA.\label{EnIntSphase}
\end{align}
The first energy functional, $W_b$, is the total bending energy of the interface
where $k_c(q)$ and $c_0(q)$ are the bending rigidity and spontaneous curvature, respectively.
The second energy functional, $W_s$, is the energy component due to splay distortion in the membrane where $k_g(q)$ is
the Gaussian bending rigidity. 
The energy due to surface tension is given by $W_\gamma$, where $\gamma$ is a
non-uniform surface tension. The coupling energy between total curvature and surface lipid phase is $W_{c}$, where $\eta$ is the coupling
constant. $W_{ext}$ is the energy due to external force field, where $\vec{F}$ is a generic
spatially varying field and $m(q)$ is a material parameter. For example, if $\vec{F}$ is
the magnetic field force then $m(q)$ would be the magnetic
susceptibility while if $\vec{F}$ is the electric field
force then $m(q)$ would be the permittivity~\cite{kolahdouz2015electrohydrodynamics}. 
The final energy, $W_{q}$, is the phase-field energy which has two contributions. 
The first is an interface energy where $k_f$ is a constant and the second is a mixing energy function $g(q)$.

Unlike prior works, material parameters such as bending rigidities and spontaneous curvatures
are taken to vary with the underlying concentration field $q$.
In the remainder of this section each energy is considered separately and for each the
expressions for $\vec{\mathcal{F}}_\Gamma$ and $\mathcal{F}_q$ are determined.

\subsection{Total Bending Energy}

The bending free energy in Eq.~(\ref{EnIntBending}) leads to
the following free energy density $w$,
\begin{align}
	w= \frac{k_c(q) (H-c_0(q))^2}{2}.
\end{align}
From this energy density the conjugate variables become
\begin{align}
	\beta&=0, \\
	\vec{\Lambda}&=\left[k_c(q)(H-c_0(q))\right]\vec{P},\\
	a&=\dfrac{k_c'(q)}{2}\left(H-c_0(q)\right)^2-k_c(q)\left(H-c_0(q)\right)c'_0(q),\\
	\vec{b} &= 0, \\
	f&=0,
\end{align}
where $c'_0(q)$ is the derivative of spontaneous curvature and $k'_c(q)$ is
derivative of bending rigidity with respect to $q$. 
Note that this free energy density does not have an explicit dependence on the position of the interface and thus $f=0$.
For this section and all after, the functional dependencies of various quantities on the concentration $q$ will be suppressed from the notation
after defining the conjugate variables.

Introduce the modified total curvature as $\tilde{H}=H-c_0$.
The tensor $\vec{\sigma}$ due to the bending energy can be computed using Eq.~\eqref{sigmatot} as follows,
\begin{align}
	\vec{\sigma}=&\frac{\tilde{H}^2 k_c}{2}\vec{P} - k_c \tilde{H}\vec{L}+\vec{n}\otimes\left[\vec{P}\nabla_s \cdot (k_c\tilde{H}\vec{P})\right]. 
	\label{eq:sigmaB}
\end{align}
The surface divergence of the first term can be expanded to
\begin{align}
	\nabla_s \cdot \left( \frac{\tilde{H}^2k_c}{2}\vec{P}\right) &=
	\frac{1}{2}\nabla_s\left(\tilde{H}^2 k_c\right)-\frac{1}{2}\tilde{H}^2 k_c H\vec{n}
	\label{bendt1}
\end{align}
while the second is
\begin{align}
	\nabla_s \cdot \left(k_c\tilde{H}\vec{L}\right) = \vec{L}\nabla_s\left(k_c\tilde{H}\right) + k_c\tilde{H}\nabla_s\cdot\vec{L}.
\end{align}
In the last term the quantity $\vec{P}\nabla_s \cdot (k_c\tilde{H}\vec{P})$ can be written as
\begin{align}
	\vec{P}\nabla_s \cdot (k_c\tilde{H}\vec{P}) 
		= \vec{P}\left[\nabla_s\left(k_c\tilde{H}\right) - k_c\tilde{H}H\vec{n}\right]
		= \vec{P}\nabla_s\left(k_c\tilde{H}\right) - k_c\tilde{H}H\vec{P}\vec{n}
		= \nabla_s\left(k_c\tilde{H}\right),
\end{align}
which leads to
\begin{align}
	\nabla_s\cdot\left\{\vec{n}\otimes\left[\vec{P}\nabla_s \cdot (k_c\tilde{H}\vec{P})\right]\right\} 
		= \nabla_s\cdot\left[\vec{n}\otimes\nabla_s\left(k_c\tilde{H}\right)\right]
		= \vec{L}\nabla_s\left(k_c\tilde{H}\right) + \vec{n}\Delta_s\left(k_c\tilde{H}\right).
\end{align}

Using these expressions the surface divergence of Eq. (\ref{eq:sigmaB}) can be written as
\begin{align}
	\nabla_s\cdot\vec{\sigma} &= \frac{1}{2}\nabla_s\left(\tilde{H}^2 k_c\right)-\frac{1}{2}\tilde{H}^2 k_c H\vec{n}
			- \vec{L}\nabla_s\left(k_c\tilde{H}\right) - k_c\tilde{H}\nabla_s\cdot\vec{L}
			+ \vec{L}\nabla_s\left(k_c\tilde{H}\right) + \vec{n}\Delta_s\left(k_c\tilde{H}\right) \nonumber \\
			&= \frac{1}{2}\nabla_s\left(\tilde{H}^2 k_c\right)-\frac{1}{2}\tilde{H}^2 k_c H\vec{n}
			 - k_c\tilde{H}\nabla_s\cdot\vec{L}
			 + \vec{n}\Delta_s\left(k_c\tilde{H}\right)
\end{align}

Using the expressions for $\tilde{H}$ and $\nabla\cdot\vec{L}$ the 
variation of the energy with respect to the interface is
\begin{align}
	\vec{\mathcal{F}}_\Gamma=-\nabla_s\cdot\vec{\sigma}=&
			-\frac{1}{2}\nabla_s\left[k_c\left(H-c_0\right)^2\right]
			+\frac{1}{2}k_c H\left(H-c_0\right)^2\vec{n} \nonumber \\
			&+ k_c\left(H-c_0\right)\left(\nabla_s H - H^2\vec{n}+2 K\vec{n}\right)
		- \vec{n}\Delta_s\left[k_c\left(H-c_0\right)\right].
	\label{eq:varTotalBending}
\end{align}
The standard Euler-Lagrange equation associated with the normal shape variation can be obtained by 
setting $\vec{\mathcal{F}}_\Gamma\cdot\vec{n}=0$ and assuming that $k_c$ and $c_0$ are constant values on the interface~\cite{Napoli2010}:
\begin{align}
	\dfrac{k_c}{2}\left(H-c_0\right)\left(H^2+c_0 H-4 K\right)+k_c\Delta_s H=0.
\end{align}

Next, consider the variation associated with the concentration field $q$. As $\vec{b}=0$ this is simply
\begin{align}
	\mathcal{F}_q=\dfrac{k_c'}{2}\left(H-c_0\right)^2-k_c\left(H-c_0\right)c'_0.
\end{align}
In the situation that material properties do not depend on the concentration field the
total bending energy has no influence on the concentration field.

\subsection{Splay Bending Energy}
The splay bending energy in Eq.~(\ref{EnIntSplay}) leads
to the following free energy density $w$,
\begin{align}
	w= k_g(q) K.
\end{align}
From this the conjugate variables are 
\begin{align}
	\beta&=0,\\
	\vec{\Lambda}&=\frac{k_g(q)}{2}\left[\frac{\partial(\tr \vec{L})^2}{\partial \vec{L}}-\frac{\partial \tr(\vec{L}^2)}{\partial \vec{L}}\right]
		=\frac{k_g(q)}{2}\left[2(\tr \vec{L})\vec{P}-2\vec{L}\right] = -k_g(q)(\vec{L}-H\vec{P}),\\
	a&= k_g'(q)K, \\
	\vec{b}&= 0, \\
	f&=0,
\end{align}
where $k'_g(q)$ is derivative of bending rigidity with respect to $q$.

The $\vec{\sigma}$ tensor due to splay
energy can be computed as
\begin{align}
	\vec{\sigma} &= k_g K\vec{P}+\vec{L}k_g(\vec{L}-H\vec{P})-\vec{n}\otimes\left\{\vec{P}\nabla_s\cdot[k_g(\vec{L}-H \vec{P})]\right\} \nonumber\\
	&= k_g\left( K\vec{P}+\vec{L}^2-H\vec{L}\right)-\vec{n}\otimes\left\{\vec{P}\nabla_s\cdot[k_g(\vec{L}-H \vec{P})]\right\}.
\end{align}
Using the Cayley-Hamilton Theorem, $\vec{L}^2 -H\vec{L}+K\vec{P}=0$, this simplifies to
\begin{align}
	\vec{\sigma} = -\vec{n}\otimes\left\{\vec{P}\nabla_s\cdot[k_g(\vec{L}-H \vec{P})]\right\}.
\end{align}
The inner expression can be evaluated as
\begin{align}
	\nabla_s\cdot[k_g(\vec{L}-H \vec{P})]
		&=\nabla_s\cdot\left(k_g\vec{L}\right)-\nabla_s\cdot\left(k_g H \vec{P}\right)\nonumber\\
		&=\vec{L}\nabla_s k_g + k_g\nabla_s\cdot\vec{L} - \nabla_s\left(k_g H\right)+k_g H^2 \vec{n} \nonumber \\
		&=\vec{L}\nabla_s k_g + k_g\left(\nabla_s H - H^2\vec{n}+2 K\vec{n}\right) - \nabla_s\left(k_g H\right)+k_g H^2 \vec{n}.
\end{align}
When including the projection operator and noting that $\nabla_s\left(k_g H\right)=k_g\nabla_s H+H \nabla_s k_g$ this becomes 
\begin{align}
	\vec{P}\nabla_s\cdot[k_g(\vec{L}-H \vec{P})] = 
		\vec{L}\nabla_s k_g - H \nabla_s k_g.
\end{align}
Thus the tensor simplifies to
\begin{align}
	\vec{\sigma} = -\vec{n}\otimes\left(\vec{L}\nabla_s k_g\right) + \vec{n}\otimes\left(H\nabla_s k_g\right).
\end{align}

The surface divergence of the first term results in
\begin{align}
	\nabla_s\cdot\left[\vec{n}\otimes\left(\vec{L}\nabla_s k_g\right)\right]
		&=\left(\nabla_s\vec{n}\right)\left(\vec{L}\nabla_s k_g\right)+\vec{n}\nabla_s\cdot\left(\vec{L}\nabla_s k_g\right) \nonumber \\
		&=\vec{L}^2\nabla_s k_g+\vec{n}\left[\left(\nabla_s k_g\right)\cdot\left(\nabla_s\cdot\vec{L}\right) + \vec{L}:\nabla_s \nabla_s k_g  \right] \nonumber \\
		&=\vec{L}^2\nabla_s k_g+\vec{n}\left(\nabla_s k_g\right)\cdot\left(\nabla_s H\right) + \vec{n}\left(\vec{L}:\nabla_s \nabla_s k_g\right).
\end{align}
The surface divergence of the second term is
\begin{align}
	\nabla_s\cdot\left[\vec{n}\otimes\left(H\nabla_s k_g\right)\right] 
		&= \left(\nabla_s\vec{n}\right)H\nabla_s k_g+\vec{n}\nabla_s\cdot\left(H\nabla_s k_g\right) \nonumber \\
		&= H\vec{L}\nabla_s k_g+\vec{n}\left[\left(\nabla_s k_g\right)\cdot\left(\nabla_s H\right)+H\Delta_s k_g\right].
\end{align}
Combining these two results with the Cayley-Hamilton Theorem the variation of 
with respect to the interface is
\begin{align}
	\mathcal{\vec{F}}_\Gamma=-\nabla_s\cdot\vec{\sigma}= -K \nabla k_g+ \vec{n}\left(\vec{L}:\nabla_s \nabla_s k_g - H\Delta_s k_g\right).
	\label{eq:varSplayBending}
\end{align}

Due to the simple nature of the conjugate variables, the variation of the energy with respect to the concentration field is simply
\begin{align}
	\mathcal{F}_q=k_g' K.
\end{align}
In the case that material properties are de-coupled from the concentration field
both $\vec{\mathcal{F}}_\Gamma$ and $\mathcal{F}_q$ are zero. The fact that $\vec{\mathcal{F}}_\Gamma=\vec{0}$ 
in this case should be expected as the Gauss-Bonnet theorem states that $\int_\Gamma K\;dA$ is
a constant for an interface with a fixed Euler characteristic. So long as the interface
has a fixed topology, the splay bending energy should not have any influence when $k_g$ is a constant.

\subsection{Tension Energy}
The tension energy leads to the following free energy density $w$,
\begin{align}
	w= \gamma(q).
\end{align}
The conjugate variables are given by
\begin{align}
	\beta&=0, \\
	\vec{\Lambda}&=0,\\
	a&= \gamma'(q),\\
	\vec{b}&= 0, \\
	f&=0,
\end{align}
where $\gamma'(q)$ is derivative of tension with respect to $q$.  

The $\vec{\sigma}$ tensor due to tension can
be computed using Eq.~(\ref{sigmatot}),
\begin{align}
	\vec{\sigma}&= \gamma \vec{P}.
\end{align}
The variation of the tension energy with respect to interface changes is
given by 
\begin{align}
	\vec{\mathcal{F}}_\Gamma=-\nabla_s\cdot\vec{\sigma}=-\nabla_s \gamma + \gamma H\vec{n}.
	\label{eq:varTension}
\end{align}
The variation of the energy with respect to the concentration field is simply
\begin{align}
	\mathcal{F}_q = a =\gamma'.
\end{align}

\subsection{Coupling Energy}
The free energy density of the coupling term is 
\begin{align}
	w= \eta H q.
\end{align}
This leads to the following conjugate variables,
\begin{align}
	\beta&=0,\\
	\vec{\Lambda}&=\eta q \vec{P},\\
	a&=\eta H,\\
	\vec{b}&= 0,\\
	f&=0.
\end{align}
The $\vec{\sigma}$ tensor due to the coupling energy is thus
\begin{align}
	\vec{\sigma} &= \eta q H\vec{P}-\eta q\vec{L}+ \vec{n}\otimes\left\{\vec{P}\nabla_s\cdot\left(\eta q \vec{P}\right)\right\} \nonumber \\
			&= \eta q H\vec{P}-\eta q\vec{L}+ \vec{n}\otimes\left\{\vec{P}\left(\nabla_s\left(\eta q\right)-\eta q H \vec{n}\right)\right\} \nonumber\\
			&= \eta q H\vec{P}-\eta q\vec{L}+ \vec{n}\otimes\nabla_s\left(\eta q\right).
\end{align}

The surface divergence of each term is given by
\begin{align}
	\nabla_s\cdot\left(\eta q H\vec{P}\right) &= \eta \nabla_s\left(q H\right) - \eta q H^2\vec{n},\\
	\nabla_s\cdot\left(\eta q\vec{L}\right) &= \eta \vec{L}\nabla_s q+\eta q \nabla_s\cdot\vec{L},\\
	\nabla_s\cdot\left[\vec{n}\otimes\nabla_s\left(\eta q\right)\right]	&= \eta \vec{L}\nabla_s q+\eta \vec{n}\Delta_s q.
\end{align}

Using these expressions the variation of the coupling energy with respect to the interface is
\begin{align}
	\vec{\mathcal{F}}_\Gamma =-\nabla_s\cdot\vec{\sigma}=& -\eta \nabla_s\left(q H\right) + \eta q H^2\vec{n}
				+ \eta \vec{L}\nabla_s q+\eta q \nabla_s\cdot\vec{L}
				-\eta \vec{L}\nabla_s q-\eta \vec{n}\Delta_s q \nonumber \\
		=&-\eta q \nabla_s H -\eta H\nabla_s q + \eta q H^2\vec{n} 
			+ \eta q \left(\nabla_s H - H^2\vec{n}+2 K\vec{n}\right) - \eta \vec{n}\Delta_s q \nonumber \\
		=&\eta\left(2 q K\vec{n} - H\nabla_s q - \vec{n}\Delta_s q\right).
	\label{eq:varCoupling}		
\end{align}

As $\vec{b}=\vec{0}$ the variation of the energy with respect to the concentration field is simply
\begin{align}
	\mathcal{F}_q&=\eta H. 
\end{align}

\subsection{External Field Energy}
The free energy density due to a generic external field is given by
\begin{align}
	w=\frac{m(q)}{2}\vec{F}(\vec{r})\cdot\vec{F}(\vec{r}),
\end{align}
where $m(q)$ is a spatially dependent material property associated with the external, spatially-varying field $\vec{F}(\vec{r})$.

The conjugate variables in this case are
\begin{align}
	\beta&=0, \\
	\vec{\Lambda}&=0,\\
	a&= \frac{m'(q)}{2}F^2(\vec{r}),\\
	\vec{b}&= 0,\\
	f&= m(q)\vec{F}(\vec{r})\cdot\nabla\vec{F}(\vec{r})\cdot\vec{n}=m(q)\vec{F}(\vec{r})\cdot\dfrac{\partial \vec{F}(\vec{r})}{\partial n},
\end{align}
where $m'(q)$ is derivative of the material parameter with respect to $q$, $F^2=\vec{F}\cdot\vec{F}$,
and $\partial \vec{F}/\partial n$ is the variation of the $\vec{F}$ field in the direction normal to the interface.

The $\vec{\sigma}$ tensor due to the external field is 
\begin{align}
	\vec{\sigma}=\frac{m}{2}F^2\vec{P}.
\end{align}
The surface divergence of this tensor is
\begin{align}
	\nabla_s\cdot\vec{\sigma}&=\frac{1}{2}\left[\nabla_s(m F^2)-mF^2H\vec{n}\right].
\end{align}
From this expression the variation of the free energy with respect to changes of the interface is
\begin{align}
	\vec{\mathcal{F}}_\Gamma=f\vec{n}-\nabla_s\cdot\vec{\sigma}=m \vec{F}\cdot\dfrac{\partial \vec{F}}{\partial n}\vec{n}-\frac{1}{2}\nabla_s(m F^2)+\dfrac{1}{2}mF^2H\vec{n}.
	\label{eq:varExternal}
\end{align}
When comparing this expression to the variation for the tension energy, Eq. (\ref{eq:varTension}), it becomes apparent 
that an external field induces a tension-like variation where $m F^2/2$ is an effective tension,
in addition to a contribution in the direction normal to the interface.

For an external field the variation of energy with respect to the concentration field is simply
\begin{align}
	\mathcal{F}_q &= \dfrac{m'}{2}F^2.
\end{align}

\subsection{Phase Energy}

The phase free energy density is
\begin{align}
	w= \frac{k_f}{2}(\|\nabla_s q\|^2)+g(q).
\end{align}
From this energy density the conjugate variables become
\begin{align}
	\beta&=0,\\
	\vec{\Lambda}&=0,\\
	a&=g'(q),\\
	\vec{b}&= k_f \nabla_sq,\\
	f&=0,
\end{align}
which defines the $\vec{\sigma}$ tensor as
\begin{align}
	\vec{\sigma} &=\frac{k_f}{2}\|\nabla_s q\|^2\vec{P} +g\vec{P} - k_f\nabla_s q \otimes \nabla_s q.
\end{align}

The surface divergence of the first term is
\begin{align}
	\nabla_s\cdot\left(\frac{k_f}{2}\|\nabla_s q\|^2\vec{P}\right) 
		&= \frac{k_f}{2}\nabla_s\left(\|\nabla_s q\|^2\right) - \frac{k_f}{2}\|\nabla_s q\|^2 H \vec{n} \nonumber \\
		&= k_f\nabla_s q\cdot\nabla_s\nabla_s q - \frac{k_f}{2}\|\nabla_s q\|^2 H \vec{n},
\end{align}
while the the surface divergence of the second term is
\begin{align}
	\nabla_s\cdot\left(g\vec{P}\right)=\nabla_s g - gH\vec{n}.
\end{align}
The final term results in
\begin{align}
	\nabla_s\cdot(k_f(\nabla_s q \otimes \nabla_s q )) 
		&= k_f\nabla_s\cdot\left(\nabla_s q \otimes \nabla_s q \right)
			= k_f \left[\left(\nabla_s\nabla_s q\right)\nabla_s q+\left(\nabla_s q\right)\Delta_s q\right].
\end{align}
From these expressions the variation of the free energy with respect to the interface is
\begin{align}
	\vec{\mathcal{F}}_\Gamma 
		&=-\nabla_s\cdot\vec{\sigma} \nonumber \\
		&= -k_f\nabla_s q\cdot\nabla_s\nabla_s q + \frac{k_f}{2}\|\nabla_s q\|^2 H \vec{n}			
			+k_f \left(\nabla_s\nabla_s q\right)\nabla_s q + k_f \left(\nabla_s q\right)\Delta_s q - \nabla_s g + gH\vec{n} \nonumber \\
		&=-k_f\left(\nabla_s q\cdot\vec{L}\nabla_s q\right)\vec{n} + \frac{k_f}{2}\|\nabla_s q\|^2 H \vec{n}			
			+k_f \left(\nabla_s q\right)\Delta_s q - \nabla_s g + gH\vec{n},
		\label{eq:varPhase}
\end{align}
where the relation $\nabla_s q\cdot\nabla_s\nabla_s q-\left(\nabla_s\nabla_s q\right)\nabla_s q=\left(\nabla_s q\cdot\vec{L}\nabla_s q\right)\vec{n}$,
as shown in Appendix \ref{sec:derivedExpressions}, has been used.

The variation of the energy with respect to changes of the concentration field is 
\begin{align}
	\mathcal{F}_q=a-\nabla_s\cdot\vec{b}_s=g' - k_f\Delta_s q,
\end{align}
which matches prior results for the chemical potential in the Cahn-Hilliard formulation.

\section{Discussion}
\label{sec:discussion}
Based on elegant framework of Napoli and Vergori, the variation of free energies associated 
with soft matter interfaces have been presented. 
These variations take into account not only the dependence of the energy on the interface 
configuration and the distribution of a surface concentration on that interface, but also 
take into account concentration-dependent material properties.

These variations can be used to determine the equilibrium shape equations for a wide
number of material systems, including but not limited to multicomponent vesicles
and surfactant-covered droplets. Due to the complex coupling between material properties,
surface concentration, and interface shape, closed-form solutions to these equilibrium 
equations would be difficult to obtain. 

If the system is not at equilibrium the derived variational expressions 
can be used to form the forces which drive the system. For example, consider a multi-component
vesicle which includes all of the energies considered here, Eqs. (\ref{EnIntBending})-(\ref{EnIntSphase}),
surrounded by a fluid. The force that the membrane exerts on the surrounding fluid, $\vec{F}_{mem}$, can calculated
by summing the contribution from each individual energy.
These forces must be balanced by changes in the fluid stress tensor, $\vec{T}$, across the interface,
$\left(\vec{T}_{out}-\vec{T}_{in}\right)\cdot \vec{n}=\vec{F}_{mem}$.
This condition can be used in conjunction with a wide number of fluid momentum solvers to obtain the influence
of the interface on the surrounding fluid. Note that due to the generalized nature of the expressions derived here
components which are normally not used in this force balance can be included. For example, it is common
to ignore the tangential contributions from the total bending variation
when modeling the dynamics of single-component lipid vesicles~\cite{salac2012,Schwalbe2011,Vlahovska2007}.
This may cause errors when considering multicomponent vesicles as tangential flow will influence the 
surface concentration field.

The variation of the energies with respect to the surface concentration can be similarly used to describe the dynamics
of the $q$ field. Using a Cahn-Hilliard model for surface phase evolution, the summation 
of each individual energy contribution would be called the chemical potential $\mu$. Using this chemical 
potential the evolution of the surface concentration field could be described by 
$\partial q/\partial t=\nabla_s\left(\nu(q) \nabla_s\mu\right)$, where $\nu(q)$ is a concentration-dependent
mobility.  The results here indicate that when additional contributions are included in the interface 
free energy, it is not sufficient to simply use standard Cahn-Hilliard models. The influence of, 
for example the total bending energy, must be taken into account when modeling complex
systems such as multicomponent vesicles.

\begin{appendices}

\section{Surface Calculus}
\label{sec:surfaceCalculus}

When considering derivatives of quantities defined on a curved surface, the variation of
the underlying surface must be taken into account. In this section
various surface vector calculus identities used to derive the energy variations are derived.
Here $a$ and $b$ be are generic scalar fields  while $\vec{u}$ and $\vec{v}$ 
are generic vectors, all defined on the interface.

\subsection{Basics}

Let the interface be orientable with an outward unit normal $\vec{n}$. 
Without loss of generality, it is assumed that the interface is described as the zero contour of 
a function $\Psi$ such that $\Psi$ is the solution to the Eikonal equation, $|\nabla\Psi|=1$
within a distance of $r$ to the interface, where $r$ depends on the curvature of the interface.
With this assumption the normal is simply $\vec{n}=\nabla\Psi$. As the normal is now defined in a small region
surrounding the interface, quantities such as the gradient of the unit normal, $\nabla\vec{n}$, are
well-defined near the interface.

The projection operator is 
given by $\vec{P}=\vec{I}-\vec{n}\otimes\vec{n}$, or in component form $P_{ij}=\delta_{ij}-n_i n_j$, where $\delta_{ij}$ 
is the Kronecker delta function. In this work, indices $i$ and $j$ are free indices and 
other indices are dummy indices. The projection operator is symmetric, $\vec{P}=\vec{P}^T$, and idempotent,
\begin{align}
	\left[\vec{P}\vec{P}\right]_{ij}
		& =P_{ip}P_{pj}   =\left(\delta_{ip}-n_i n_p\right)\left(\delta_{pj}-n_p n_j\right)  =\delta_{ip}\delta_{pj}-n_i n_p \delta_{pj} - n_p n_j \delta_{ip} + n_i n_p n_p n_j  \nonumber \\
		& = \delta_{ij} - n_i n_j - n_i n_j + n_i n_j  = \delta_{ij} - n_i n_j = \left[\vec{P}\right]_{ij},
\end{align}
where $\left[\vec{v}\right]_i$ is the $i$-component of a vector $\vec{v}$, $\left[\vec{A}\right]_{ij}$ is the 
$i,j$-component of a tensor $\vec{A}$, and repeated indices indicate summation. There are also no components of $\vec{P}$ in
the normal direction,
\begin{align}
	[\vec{P}\vec{n}]_i =P_{ip}n_{p}&=(\delta_{ip}-n_in_p)n_p=\delta_{ip}n_p-n_in_p n_p=n_i-n_i =0. \label{pn0}
\end{align}

The generalized surface gradient function can be written as
$\nabla_s\vec{A}=\left(\nabla\vec{A}\right)\cdot\vec{P}$, where $\vec{A}$ can be either a scalar, vector, or tensor field~\cite{Fried2008,Gurtin1975,Napoli2012}.
For example, the surface gradient of a scalar field $a$ in component form would be written as 
\begin{align}
	\left[\nabla_s a\right]_i
		=\left[\left(\nabla a\right)\cdot\vec{P}\right]_i
		=\dfrac{\partial a}{\partial x_p}P_{pi},
\end{align}
while the surface gradient for a vector field $\vec{v}$ would be
\begin{align}
	\left[\nabla_s \vec{v}\right]_{ij}
		=\left[\left(\nabla \vec{v}\right)\cdot\vec{P}\right]_{ij}
		=\dfrac{\partial v_i}{\partial x_p}P_{pj}.
\end{align}

The surface divergence of any vector $\vec{v}$ can be written as 
$\nabla_s\cdot\vec{v}=\tr\nabla_s\vec{v}=\vec{P}:\nabla\vec{v}$~\cite{Fried2008}.
In component form this is written as
\begin{align}
	\left[\nabla_s\cdot\vec{v}\right]
		=\left[\vec{P}:\nabla\vec{v}\right]
		=\dfrac{\partial v_p}{\partial x_q}P_{pq}.
\end{align}
The surface divergence of a tensor field $\vec{A}$ is defined as~\cite{Fried2008}
\begin{align}
	\left[\nabla_s\cdot\vec{A}\right]_{i}=\left[\left(\nabla \vec{A}\right):\vec{P}\right]_i=\dfrac{\partial A_{ip}}{\partial x_q}P_{pq}.
\end{align}


\subsection{Derivatives of Interface Quantities}
The second fundamental form of the deformable interface $\Gamma$ is $\vec{L}=\nabla_s \vec{n}$, which is
a second order symmetric tensor tangential to the interface and describes the interface curvatures via the first and second invariants:
\begin{align}
	H=\nabla_s\cdot\vec{n}=\tr \vec{L} \quad \textnormal{and} \quad K=\frac{1}{2}\left[(\tr\vec{L})^2-\tr(\vec{L}^2)\right].
\end{align}
The tensor $\vec{L}$ has no components in the normal direction,
\begin{align}
	[\vec{L}\cdot\vec{n}]_i&=[(\nabla_s\vec{n})\cdot\vec{n}]_i=\frac{\partial n_i}{\partial x_q}P_{qp} n_p=0,\label{ln0}
\end{align}
and does not change under application of $\vec{P}$:
\begin{align}
	[\vec{L}\vec{P}]_{ij}=[\nabla_s \vec{n}\vec{P}]_{ij}=\frac{\partial n_i}{\partial x_p} P_{pq}P_{qj}=\frac{\partial n_i}{\partial x_p}P_{pj}
		=[\nabla_s\vec{n}]_{ij} =[\vec{L}]_{ij}, \label{lp}
\end{align}
and
\begin{align}
	\vec{L}=\vec{L}^T=\left(\vec{L}\vec{P}\right)^T=\vec{P}^T\vec{L}^T=\vec{P}\vec{L}.
\end{align}
The full contraction of the projection operator and $\vec{L}$ results in
\begin{align}
	\left[\vec{P}:\vec{L}\right]&=P_{pq}L_{pq}=(\delta_{pq} -n_pn_q)\left(\frac{\partial n_p}{\partial x_r}P_{rq}\right)
		=\delta_{pq}\left(\frac{\partial n_p}{\partial x_r}P_{rq}\right) -n_p\frac{\partial n_p}{\partial x_r}P_{rq}n_q \nonumber \\
		&=\frac{\partial n_p}{\partial x_r}P_{rq} =\left[\nabla_s\cdot \vec{n}\right] = \left[H\right]\label{PL}
\end{align}

The surface divergence of the projection operator and of the second fundamental form have been provided
previously~\cite{Napoli2010}. After taking into account the definition of the curvature used here they are,
\begin{align}
	\nabla_s\cdot\vec{P}&=-H\vec{n} \label{eq:sdivP}\\
	\nabla_s\cdot\vec{L} &= \nabla_s H - H^2\vec{n}+2 K\vec{n}. \label{eq:sdivL}
\end{align}

\subsection{Derived Expressions}
\label{sec:derivedExpressions}

The surface gradient of the multiple of two scalar fields is
\begin{align}		
	\left[\nabla_s \left(a b\right)\right]_i
		&=\left[\left(\nabla a b\right)\cdot\vec{P}\right]_i=\dfrac{\partial \left(a b\right)}{\partial x_p}P_{pi} 
			=\left(a\dfrac{\partial b}{\partial x_p}+b\dfrac{\partial a}{\partial x_p}\right)P_{pi}=\left[a \nabla_s b+b \nabla_s a\right]_i
\end{align}
while the surface gradient of a vector dot product is
	\begin{align}
		\left[\nabla_s\left(\vec{u}\cdot\vec{v}\right)\right]_i
			&=\left[\nabla \left(\vec{u}\cdot\vec{v}\right)\cdot\vec{P}\right]_i
				=\dfrac{\partial\left(u_p v_p\right)}{\partial x_q}P_{qi}\nonumber \\
			&= v_p \dfrac{\partial u_p}{\partial x_q}P_{qi} + u_p\dfrac{\partial v_p}{x_q}P_{qi} \nonumber \\
			&= \left[\vec{v}\cdot\nabla_s\vec{u}+\vec{u}\cdot\nabla_s\vec{v}\right]_i.
	\end{align}

The surface divergence of a scalar field $a$ times a vector $\vec{u}$ is
\begin{align}
	\left[\nabla_s\cdot\left(a\vec{u}\right)\right]
		=\dfrac{\partial\left(a u_p\right)}{\partial x_q}P_{pq}
		=\dfrac{\partial a}{\partial x_q} u_p P_{pq}+\dfrac{\partial u_p}{\partial x_q} a P_{pq}
		=\left[\vec{u}\cdot\nabla_s a+a\nabla_s\cdot\vec{u}\right].
\end{align}

The surface divergence of a scalar field times a surface tensor is
\begin{align}
	[\nabla_s\cdot(a \vec{A})]_i&=[\nabla(a \vec{A})\cdot\vec{P}]_i
			=\frac{\partial(a A_{ip})}{\partial x_q}P_{qp}
			=\frac{\partial a}{\partial x_q}A_{ip}P_{qp}+ a\frac{\partial A_{ip}}{\partial x_q}P_{qp}\nonumber \\
		&=\left[\vec{A}\nabla_s a+a \nabla_s\cdot \vec{A}\right]_i.
\end{align}
When the surface tensor is the projection operator this results in
\begin{align}
	\nabla_s\cdot(a \vec{P}) &=\vec{P}\nabla_s a+a \nabla_s\cdot \vec{P} = \nabla_s a - a H\vec{n} \label{sdiv_ap},
\end{align}
while when the surface tensor is $\vec{L}$ then
\begin{align}
	\nabla_s\cdot(a \vec{L}) &=\vec{L}\nabla_s a+a \nabla_s\cdot \vec{L} = \vec{L}\nabla_s a + a \left(\nabla_s H - H^2\vec{n}+2 K\vec{n}\right) \label{sdiv_al}.
\end{align}

The surface divergence of the tensor (outer) product between two vectors is given by
\begin{align}
	\left[\nabla_s\cdot\left(\vec{u}\otimes\vec{v}\right)\right]_i
		& =\left[\left\{\nabla\left(\vec{u}\otimes\vec{v}\right)\right\}\cdot\vec{P}\right]_i 
				=\dfrac{\partial\left(u_i v_p\right)}{\partial x_q}P_{qp} \nonumber \\
		& = \dfrac{\partial u_i}{\partial x_q}P_{qp}v_p +u_i \dfrac{\partial v_p}{\partial x_q}P_{pq} 
				= \left[\left(\nabla_s\vec{u}\right)\vec{v} + \vec{u}\nabla_s\cdot\vec{v}\right]_i
\end{align}

The surface divergence of a symmetric tensor $\vec{A}$ times a vector $\vec{u}$ is
\begin{align}
	\left[\nabla_s\cdot\left(\vec{A}\vec{u}\right)\right]
		&= P_{pq}\dfrac{\partial}{\partial x_q}\left(A_{pr}u_r\right) 
			= P_{pq}\dfrac{\partial A_{pr}}{\partial x_q}u_r + P_{pq}A_{pr}\dfrac{\partial u_r}{\partial x_q} \nonumber\\
		&= u_r\dfrac{\partial A_{rp}}{\partial x_q}P_{pq} + A_{rp}\dfrac{\partial u_r}{\partial x_q}P_{pq} 
			= \left[\vec{u}\cdot\left(\nabla_s\cdot\vec{A}\right)+\vec{A}:\nabla_s\vec{u}\right].
\end{align}

Finally, consider the simplification of $\nabla_s a\cdot\nabla_s\nabla_s a-\left(\nabla_s\nabla_s a\right)\nabla_s a$.
First, the surface Hessian of a scalar field $a$ is
\begin{align}
	\left[\nabla_s\nabla_s a\right]_{ij}
		=\dfrac{\partial}{\partial x_p}\left(\dfrac{\partial a}{\partial x_q}P_{qi}\right)P_{pj}
		=\dfrac{\partial^2 a}{\partial x_p \partial x_q}P_{qi}P_{pj} + \dfrac{\partial a}{\partial x_q}\dfrac{\partial P_{qi}}{\partial x_p}P_{pj}
\end{align}
Thus,
\begin{align}
	\left[\nabla_s a\cdot\nabla_s\nabla_s a-\left(\nabla_s\nabla_s a\right)\nabla_s a\right]_i
		=&\dfrac{\partial a}{\partial x_r}P_{rs}\left(\dfrac{\partial^2 a}{\partial x_p \partial x_q}P_{qs}P_{pi} + \dfrac{\partial a}{\partial x_q}\dfrac{\partial P_{qs}}{\partial x_p}P_{pi}\right) \nonumber\\
		&-\left(\dfrac{\partial^2 a}{\partial x_p \partial x_q}P_{qi}P_{ps} + \dfrac{\partial a}{\partial x_q}\dfrac{\partial P_{qi}}{\partial x_p}P_{ps}\right)\dfrac{\partial a}{\partial x_r}P_{rs} \nonumber\\
		=&\dfrac{\partial a}{\partial x_q}\left(\dfrac{\partial P_{qs}}{\partial x_p}P_{pi}-\dfrac{\partial P_{qi}}{\partial x_p}P_{ps}\right)\dfrac{\partial a}{\partial x_r}P_{rs} \nonumber \\
		=&\dfrac{\partial a}{\partial x_q}\left(
					-n_q\dfrac{\partial n_s}{\partial x_p}P_{pi} - n_s\dfrac{\partial n_q}{\partial x_p}P_{pi}
					+n_q\dfrac{\partial n_i}{\partial x_p}P_{ps} + n_i\dfrac{\partial n_q}{\partial x_p}P_{ps}
					\right)\dfrac{\partial a}{\partial x_r}P_{rs}.		
\end{align}
Recalling that $\left[\vec{P}\cdot\vec{n}\right]_i=P_{rs}n_s=0$, 
$\left[\vec{L}\right]_{is}=\left(\partial n_i/\partial x_p\right)P_{ps}=\left(\partial n_s/\partial x_p\right)P_{pi}=\left[\vec{L}\right]_{si}$ due to the symmetry of $\vec{L}$,
and $\vec{L}=\vec{P}\vec{L}$
this expression simplifies to
\begin{align}
	\left[\nabla_s a\cdot\nabla_s\nabla_s a-\left(\nabla_s\nabla_s a\right)\nabla_s a\right]_i		
		=&\dfrac{\partial a}{\partial x_q} L_{qs}\dfrac{\partial a}{\partial x_r}P_{rs}n_i
			=\dfrac{\partial a}{\partial x_q} P_{qp}L_{ps}\dfrac{\partial a}{\partial x_r}P_{rs}n_i \nonumber \\
		=&\left[\left(\nabla_s a\cdot\vec{L}\nabla_s a\right)\vec{n} \right]_i
\end{align}

\section{First Variation of \textit{L} }
\label{sec:variationL}

The first variation of $\vec{L}$ can be obtain by using $\vec{L}=\nabla\vec{n}\vec{P}$ and the 
product rule~\cite{Napoli2010},
\begin{align}
	\delta\vec{L}=\delta(\nabla\vec{n}\vec{P})=\delta(\nabla \vec{n})\vec{P} +(\nabla \vec{n})\delta \vec{P}.
\end{align}
Variation of the gradient of normal and the projection operator are given as~\cite{Napoli2010},
\begin{align}
	\delta(\nabla \vec{n})&=-\nabla[(\nabla_s\vec{u})^T\vec{n}]-(\nabla	\vec{n})(\nabla \vec{u}),\label{var_gradn}\\
	\delta \vec{P} &= (\nabla_s\vec{u})^T\vec{n}\otimes\vec{n} +	\vec{n}\otimes(\nabla_s\vec{u})^T\vec{n}, \label{var_p}
\end{align}
where $\vec{u}$ is a virtual displacement field.
Therefore, 
\begin{align}
	\delta(\nabla\vec{n}\vec{P})&=-\nabla[(\nabla_s
	\vec{u})^T\vec{n}]\vec{P}-(\nabla \vec{n})(\nabla \vec{u})\vec{P} +(\nabla
	\vec{n})\left[(\nabla_s \vec{u})^T\vec{n}\otimes\vec{n}\right]
	+\left(\nabla\vec{n}\right)\left[\vec{n}\otimes(\nabla_s \vec{u})^T\vec{n}\right]. 
\end{align}

This expression can be simplified by noting that 
\begin{align}
	[\nabla\vec{n}(\nabla_s \vec{u})^T\vec{n}\otimes\vec{n}]_{ij} &= \frac{\partial n_i}{\partial x_s}\frac{\partial u_r}{\partial x_t}P_{ts}n_rn_j
			=\frac{\partial n_i}{\partial x_s}\frac{\partial u_r}{\partial x_t}P_{tv}P_{vs}n_rn_j\nonumber\\
		&=\frac{\partial n_i}{\partial x_s}P_{vs}\frac{\partial u_r}{\partial x_t}P_{tv}n_rn_j
		=[(\nabla_s \vec{n})(\nabla_s \vec{u})^T(\vec{n} \otimes \vec{n})]_{ij}. \label{var_gradn_interm1}
\end{align}
Additionally,
\begin{align}
	[-(\nabla \vec{n})(\nabla \vec{u})\vec{P}+(\nabla \vec{n})(\vec{n} \otimes	(\nabla_s \vec{u})^T\vec{n})]_{ij}
		=&-\frac{\partial n_i}{\partial x_p}\frac{\partial u_p}{\partial x_q}P_{qj}+\frac{\partial n_i}{\partial x_p}n_p \frac{\partial u_t}{\partial x_s}P_{sj}n_t\nonumber\\
		=&-\frac{\partial n_i}{\partial x_p}\frac{\partial u_p}{\partial x_q}\delta_{qj}+\frac{\partial n_i}{\partial x_p}\frac{\partial u_p}{\partial x_q}n_qn_j\nonumber \\ 
			&+\frac{\partial n_i}{\partial	x_p}\frac{\partial u_t}{\partial x_s}n_p n_t \delta_{sj}-\frac{\partial n_i}{\partial x_p}\frac{\partial u_t}{\partial x_s}n_p n_s n_j n_t. \label{var_gradn_interm_t0}
\end{align}
These terms can be rewritten as
\begin{align}
	\frac{\partial n_i}{\partial x_p}\frac{\partial u_p}{\partial x_q}\delta_{qj}
		&=\frac{\partial n_i}{\partial x_r}\frac{\partial u_p}{\partial x_q}\delta_{rp}\delta_{qj}
		=\frac{\partial n_i}{\partial x_r}\frac{\partial u_p}{\partial x_s}\delta_{rp}\delta_{sj},\label{var_gradn_interm_t1} \\
	\frac{\partial n_i}{\partial x_p}\frac{\partial u_p}{\partial x_q}n_qn_j
		&=\frac{\partial n_i}{\partial x_r}\frac{\partial u_p}{\partial x_q}\delta_{rp}n_qn_j
		=\frac{\partial n_i}{\partial x_r}\frac{\partial u_p}{\partial x_s}\delta_{rp}n_sn_j,\label{var_gradn_interm_t3}\\
	\frac{\partial n_i}{\partial x_p}\frac{\partial u_t}{\partial x_s}n_pn_t\delta_{sj}
		&=\frac{\partial n_i}{\partial x_r}\frac{\partial u_p}{\partial x_s}n_rn_p\delta_{sj},\label{var_gradn_interm_t2} \\
	\frac{\partial n_i}{\partial x_p}\frac{\partial u_t}{\partial x_s}n_pn_sn_jn_t
		&=\frac{\partial n_i}{\partial x_r}\frac{\partial u_t}{\partial x_s}n_rn_sn_jn_t
		=\frac{\partial n_i}{\partial x_r}\frac{\partial u_p}{\partial x_s}n_rn_sn_jn_p.\label{var_gradn_interm_t4}
\end{align}
Using these expression Eq. (\ref{var_gradn_interm_t0}) is now
\begin{align}
	[-(\nabla \vec{n})(\nabla \vec{u})\vec{P}+(\nabla \vec{n})(\vec{n} \otimes (\nabla_s \vec{u})^T\vec{n})]_{ij}
		=&-\frac{\partial n_i}{\partial x_r}\frac{\partial u_p}{\partial x_s}\delta_{rp}\delta_{sj}+\frac{\partial n_i}{\partial x_r}\frac{\partial u_p}{\partial x_s}\delta_{rp}n_sn_j  \nonumber \\
		 &+ \frac{\partial n_i}{\partial x_r}\frac{\partial u_p}{\partial x_s}n_rn_p\delta_{sj} - \frac{\partial n_i}{\partial x_r}\frac{\partial	u_p}{\partial x_s}n_r n_s n_j n_p \nonumber\\
		=&-\frac{\partial n_i}{\partial x_r}\frac{\partial u_p}{\partial x_s}\left(\delta_{rp}\delta_{sj}-\delta_{rp}n_sn_j-n_rn_p\delta_{sj}+ n_r n_s n_j n_p	\right)\nonumber\\
		=&-\frac{\partial n_i}{\partial x_r}\frac{\partial u_p}{\partial x_s}\left(\delta_{rp}(\delta_{sj}-n_sn_j)-n_rn_p(\delta_{sj}-n_sn_j) \right)\nonumber\\
		=&-\frac{\partial n_i}{\partial x_r}\frac{\partial u_p}{\partial x_s} (\delta_{rp}-n_rn_p)(\delta_{sj}-n_sn_j)\nonumber\\
		=&-\frac{\partial n_i}{\partial x_r}\frac{\partial u_p}{\partial x_s}P_{rp}P_{sj}\nonumber\\
		=&-\frac{\partial n_i}{\partial x_r}P_{rp}\frac{\partial u_p}{\partial x_s}P_{sj}\nonumber\\
		=&-[\nabla_s\vec{n}\nabla_s\vec{u}]_{ij}.
\label{var_gradn_interm2}
\end{align}

Using Eq.~(\ref{var_gradn_interm1}) and Eq.~(\ref{var_gradn_interm2}) the first variation of $\vec{L}$ is obtained as
\begin{align}
	\delta\vec{L}=\delta(\nabla \vec{n}\vec{P})&=(\nabla_s \vec{n})(\nabla_s\vec{u})^T\vec{n}\otimes\vec{n}-\nabla_s[(\nabla_s \vec{u})^T\vec{n}]-(\nabla_s \vec{n})(\nabla_s \vec{u}) \nonumber \\
			&=\vec{L}(\nabla_s\vec{u})^T\vec{n}\otimes\vec{n}-\nabla_s[(\nabla_s\vec{u}^T)\vec{n}]-\vec{L}(\nabla_s \vec{u}).
\end{align}

\end{appendices}

\begin{acknowledgement}
	This work has been supported by the National Science Foundation through the Division of Chemical, Bioengineering, Environmental, and Transport Systems Grant \#1253739.
\end{acknowledgement}


\end{document}